\documentclass{article}

\usepackage{arxiv}

\usepackage[utf8]{inputenc} 
\usepackage[T1]{fontenc}    
\usepackage{hyperref}       
\usepackage{url}            
\usepackage{booktabs}       
\usepackage{amsfonts}       
\usepackage{nicefrac}       
\usepackage{microtype}      
\usepackage{graphicx}
\usepackage[compress,square,numbers]{natbib}
\usepackage{doi}
\title{DeepWL: Robust EPID based Winston-Lutz Analysis using Deep Learning and Synthetic Image Generation}

\author{ \href{https://orcid.org/0000-0002-3862-2644}{\includegraphics[scale=0.06]{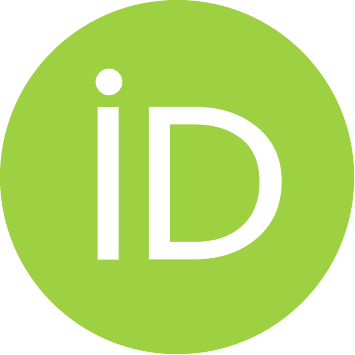}\hspace{1mm}Michael J. J. Douglass} \\
	School of Physical Sciences\\
	University of Adelaide\\
	Adelaide, South Australia, 5005 \\
	\texttt{michael.douglass@adelaide.edu.au} \\
	\And
	{James A. Keal} \\
	School of Physical Sciences\\
	University of Adelaide\\
	Adelaide, South Australia, 5005 \\
}

\date{July 2, 2021}

\renewcommand{\headeright}{Preprint}
\renewcommand{\undertitle}{Preprint: Submitted to Physica Medica}
\renewcommand{\shorttitle}{Automated Linear Accelerator QA using Deep Learning}

\hypersetup{
pdftitle={DeepWLMDouglass},
pdfsubject={q-bio.NC, q-bio.QM},
pdfauthor={Michael J. J.~Douglass, James A.~Keal},
pdfkeywords={deep learning, radiation oncology, medical physics, QA, EPID, quality assurance, linear accelerator, synthetic data, blender, machine learning, optical path tracing}
}

\begin{document}
\maketitle

\begin{abstract}
Radiation therapy requires clinical linear accelerators to be mechanically and dosimetrically calibrated to a high standard. One important quality assurance test is the Winston-Lutz test which localizes the radiation isocentre of the linac. 

In the current work we demonstrate a novel method of analysing EPID based Winston-Lutz QA images using a deep learning model trained only on synthetic image data.

In addition, we propose a novel method of generating the synthetic WL images and associated ‘ground-truth’ masks using an optical ray-tracing engine to ‘fake’ mega-voltage EPID images.

The model called DeepWL was trained on 1500 synthetic WL images using data augmentation techniques for 180 epochs. The model was built using Keras with a TensorFlow backend on an Intel Core i5-6500T CPU and trained in approximately 15 hours. 

DeepWL was shown to produce ball bearing and multi-leaf collimator field segmentations with a mean dice coefficient of 0.964 and 0.994 respectively on previously unseen synthetic testing data. When DeepWL was applied to WL data measured on an EPID, the predicted mean displacements were shown to be statistically similar to the Canny Edge detection method. However, the DeepWL predictions for the ball bearing locations were shown to correlate better with manual annotations compared with the Canny edge detection algorithm. 

DeepWL was demonstrated to analyse Winston-Lutz images with accuracy suitable for routine linac quality assurance with some statistical evidence that it may outperform Canny Edge detection methods in terms of segmentation robustness and the resultant displacement predictions.

\end{abstract}


\section{Introduction}
Radiation therapy relies on the clinical linear accelerator (Linac) being mechanically and dosimetrically calibrated to a high accuracy and regularly tested as part of a quality assurance program. The Winston-Lutz (WL) \cite{RN39} test is a procedure commonly used by medical physicists to measure the congruence between the radiation isocentre of a linac and the alignment of the treatment lasers \cite{RN35,RN36,RN37,RN38}. The WL test is also useful for diagnosing mechanical issues with the Linac such as: room laser alignment, mechanical isocentre alignment, and if the mutli-leaf collimators (MLCs) are used to define the field irradiating the test phantom, the test can also provide information about the alignment of the MLC leaves. 

\begin{figure}
	\centering
	\includegraphics[width=8cm]{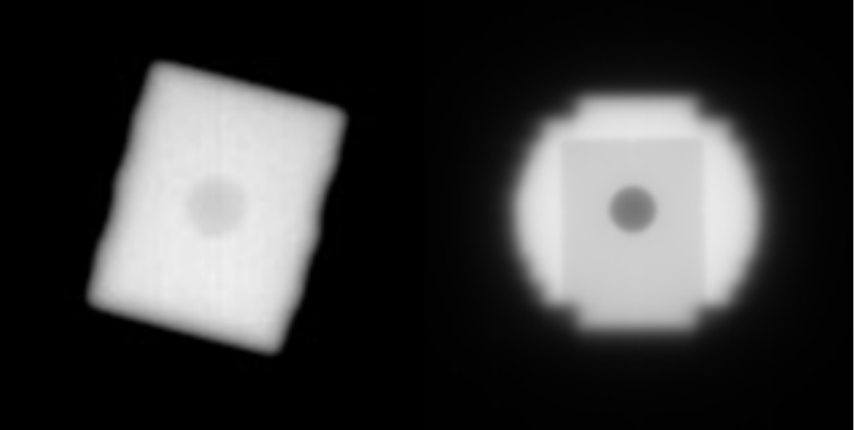}
	\caption{Two Winston-Lutz fields measured on a Varian© AS1200 EPID panel. The bright in-field region is defined by the MLCs. The dark circular area is the ball bearing. (Left) – Custom QA phantom. (Right) Brainlab SRS WL Front Pointer.}
	\label{fig:fig1}
\end{figure}

To quantitatively assess the alignment of the radiation isocentre to the treatment lasers in our centre, an ‘in-house’ designed quality assurance (QA) phantom was developed which contains a ceramic ball bearing (BB) at the centre of a plastic phantom. During routine QA, this phantom is aligned visually to the isocentre indicated by the room lasers and a series of images are acquired using the electronic portal imaging device (EPID) and a small MLC defined radiation field. In our centre we use a 1.5 cm $\times$ 1 cm MLC defined field (size defined at machine isocentre). These fields are delivered from each cardinal gantry angle, collimator angle as well as several couch angles. The resultant images (example shown on the left of figure \ref{fig:fig1}) contain a rectangular high dose region defined by the MLC leaves and a smaller dark circular region representing the radio-opaque BB. The congruence of the centre of the BB and radiation field indicate the relative alignment of the radiation isocentre to the room lasers from that gantry or collimator projection. 

There are several commercial and open source solutions available for auto-analysis of quality assurance images \cite{RN40, RN41}, however these solutions are typically only compatible with a few standard commercial QA phantoms and are not suitable for our custom QA phantom. One reason for this is the reduced contrast of the BB in the EPID image compared with the background radiation field. Our custom QA phantom was designed to be used for several routine QA processes simultaneously to improve efficiency, but the resultant QA images have reduced contrast compared with a commercial WL phantom as illustrated in the right of Figure \ref{fig:fig1}. As a result, we developed a custom python \cite{RN42} script to quantitatively analyse each of these QA images, reporting the alignment of the radiation isocentre in three dimensions relative to the lasers. This is achieved using traditional edge detection methods such as Canny edge detection (CED) \cite{RN48} and a series of morphological image processing steps to isolate the edges of the MLC field from the BB. 

The CED algorithm often produces many artefacts which must be removed through post-processing techniques. In extreme cases, particularly if the BB is close to the edge of the MLC field, the CED algorithm can group the out-of-field and BB regions into one, resulting in highly inaccurate reported results. Although, in these rare cases, the cause is typically setup error, the quantitative results that the code reports are inaccurate and may require the code to be debugged. Additionally, since our centre uses several linac models with several different EPID designs and WL QA phantoms, the threshold parameters used in the CED based WL analysis code depends on the combination of equipment used. Practically, this method of analysis can be time consuming when manual human intervention is required to interpret and resolve issues.

Machine learning, and in particular, deep learning (DL) has been adopted in numerous fields in the last 10 years. One field where DL has been rapidly adopted is medicine, with applications in diagnosis, prognosis, and treatment in many specialties. In radiation oncology, an area of particular interest is organ segmentation \cite{RN62, RN59, RN55, RN58, RN57, RN56, RN53, RN61, RN60, RN54, RN52}. One of the limitations of a rapid treatment planning process is the time required to contour/segment each organ at risk (OAR) and targets for plan optimization, and DL may reduce this burden. In addition, DL segmentation may reduce inter-observer variability when contouring a patient thus improving standardization of care between patients.

A watershed moment in DL-based image segmentation occurred in 2015, when Ronneberger et. al. \cite{RN1}  reported on their “U-Net” model. The network owes its name to the “U” shaped architecture consisting of a “contraction” and “expansion” path. The contraction path consists of a series of repeated convolutional layers and pooling layers which reduces the spatial information while increasing the number of features. The expansion path combines these features with high resolution spatial information from concatenation and deconvolution layers. The result is a pixel-wise segmentation of the original image with the same resolution as the input image, or in some cases, a cropped image. 

Since the inception of U-Net, many variants have been developed for other applications and other domains such as V-Net \cite{RN2} and DeepMedic \cite{RN3}. The high relative training and evaluation speeds on modern graphics processing units (GPUs), the small amount of data required to train (e.g. CT and contoured structures) and the simplicity of the architecture have made many segmentation tasks more tractable and provides a well optimized starting point for radiation oncology departments with limited experience to enter this domain.  Machine learning techniques have also been applied in radiation oncology for other tasks such as dose calculation \cite{RN34, RN33, RN32}, patient specific QA \cite{RN26, RN30, RN27, RN29} and Linac quality assurance \cite{RN31, RN28}.

One major obstacle to generating high accuracy DL segmentation models is the need for high quality curated training data, in the form of CT images, with carefully contoured organ structures for all slices. This generally requires one or more clinicians to manually annotate organ contours slice-by-slice on a CT or MRI dataset for several hundred patient scans. One method to reduce this burden is data augmentation techniques \cite{RN64} such as introducing horizontal or vertical flips, scaling, translations, or other variations in the training data to artificially increase the size of the training dataset. Semi-automated contouring technologies such as atlas-based segmentation \cite{RN63} on the original CT data, may provide a starting point for the oncologist during the data curation stage and reduce the total time required. Both processes have limitations and are subject to uncertainties introduced by either human subjectivity \cite{RN62}, annotation error, and in the case of an automated process, the choice of parameters in the software which may affect annotation accuracy. 

Recent studies in multiple domains have reduced this burden by utilizing synthetic training data as a means of training DL models where insufficient high-quality “real” data is unavailable. This technique is particularly useful in cases where manual annotation of the training data is too time consuming or would lead to high uncertainties such as inter-observer variability in the case of organ-at-risk segmentation. Previous examples of fields utilizing this technique include: geology \cite{RN11, RN12}, biology \cite{RN9, RN10, RN4, RN8, RN7}, automotive \cite{RN5, RN6}, medical \cite{RN16, RN15, RN13, RN20, RN18, RN17, RN14} and robotics \cite{RN23, RN22, RN24}. This technique has also been utilized in radiation oncology to calculate MV linac radiation doses in real patient CTs using DL models which were trained only on synthetic data \cite{RN25}.

In the current work, we expand on these approaches by training a DL model on synthetic images to analyse measured EPID based WL fields without transfer learning techniques. In addition, we propose a novel method of rapidly generating synthetic megavoltage (MV) WL images, using an optical path-tracing engine used typically for computer generated graphics, thus eliminating the need for complex Monte Carlo radiation transport simulations in the current work. 

\section{Aim}

The purpose of the current work was to investigate and develop a more robust algorithm for automatically and quantitatively assessing routine linac QA images measured on an EPID using DL techniques, specifically, WL images. We hypothesized that the DL model could be trained using only synthetic WL image data and generalize sufficiently to analyse “real” measured WL images without transfer learning techniques. The DL algorithm proposed in the current work utilized a U-Net architecture to produce a labelled mask of BB and radiation field from the original WL EPID image. The displacement between the centre of masses of these two masks would then be calculated for each WL field, enabling the position of the radiation isocentre to be localized relative to the room lasers.

We investigated whether a DL model trained entirely on synthetic WL images is sufficiently general to analyse measured WL images, with enough robustness and accuracy for routine quality assurance. It is hypothesized that provided there is sufficient variation in the synthetic images including: variable image scale, variable BB size, variable phantom scatter, random BB displacement and collimator rotation, the DL model will be able to robustly analyse “real” measured WL images for routine quality assurance. Based on this assumption, we also tested the theory that this synthetic data could be produced through optical path-tracing techniques typically used for computer generated renderings, rather than conventional radiation transport Monte Carlo simulations. This approach requires very little knowledge of coding, does not require precise geometric modelling, or any knowledge of the material composition of the linac components. 

\section{Materials and Methods}

\subsection{Generation of Synthetic Winston-Lutz Data using Blender}

The task of generating synthetic MV EPID images would generally require traditional Monte Carlo radiation transport algorithms to accurately simulate the x-ray and electron particle interactions between the linac target and the EPID. Such simulations, whilst producing results with high quantitative accuracy, are time consuming to develop, test and typically require many simulated histories to achieve low statistical noise in the data. This can potentially require a large amount of time to produce each simulated image. 

For the task proposed in the current work, a suitable Monte Carlo simulation would require, at minimum, modelling of the MLC leaf geometry and materials, collimator design, EPID geometry and materials to achieve a realistic response to x-rays, geometry and materials of the QA phantom, a suitable x-ray spectra, and angular distribution from the target. Much of this data is not publicly available due to confidentiality. In addition, a suitable physics list and scoring code would need to be written and validated before proceeding with generating the synthetic data. Further challenges would arise when attempting to produce corresponding synthetic pixel-wise masks corresponding to the radiation field and BB. 

Blender \cite{RN43} is an open-source 3D modelling and rendering software used typically for computer generated art and animation. Blender uses an optical path-tracing render engine called “Cycles” to render photo realistic 3D scenes containing objects with various materials. These materials have optical properties such as transmission, roughness, specularity and index of refraction, each of which can be defined using a fully customizable node-based system.

To avoid the challenges associated with commissioning a simulated linac using traditional Monte Carlo methods, Blender version 2.82 was used to build a simple Varian (Palo Alto, California) linac model, based on the approximate specifications of our linacs, using data from our Raystation (Raysearch Laboratories, Stockholm, Sweden) TPS. We propose that since the DL model will learn the “appearance” of a WL image from a large sample of randomized synthetic images, the high accuracy quantitative data achievable using traditional Monte Carlo methods is un-necessary for this application. 

The only data required to build the simulation, apart from a rudimentary understanding of the design of a linac head, was the source to EPID distance, the source to MLC distance, MLC leaf widths and the geometric dimensions of the QA phantom. Material compositions were not required for the simulation as the visual appearance of the synthetic WL images were optimised though changes in the optical properties of the materials and changes to settings in the optical path-tracing engine. These modifications enabled the synthetization of the approximate radiographic appearance of a MV image.

The optical and geometric parameters of this simulation were controlled via a custom python script designed to produce a variety of synthetic WL images of varying appearance, sufficient to encompass the radiographic appearance of WL images typically acquired from a real Linac. Providing sufficient variation is included in the synthetic training images, we hypothesize that the resultant DL model should generalize sufficiently to real WL images to produce predictions comparable to traditional algorithms. This is the first time, to our knowledge, that this approach has been used, where optical path-tracing has been used to simulate images in the MV x-ray spectrum for the purposes of DL model training.

In addition to the advantage of reduced model commissioning time, this technique allows labelled masks of the target object to be synthesised at the same time as the WL image. This results in a 1:1 correspondence between the input WL image and the output training masks, with an accuracy limited only by the selected rendering resolution. 

\subsection{Simulation Geometry}

In the Blender simulation, the x-ray source was defined as an optical point light with a variable radius and intensity. Light from this source was collected by a “camera” object in Blender which acted as a virtual EPID panel. This camera was placed 150 cm from the centre of the light source to simulate an EPID panel at 50 cm from the radiation isocentre of a Varian (Palo Alto, California) linac. The lateral position of the camera was randomized by up to 5 mm from the radiation central axis via the script to simulate lateral movements of the EPID relative to the MLC field and BB. The camera was set to orthographic mode (instead of perspective) to remove any geometric perspective effects in the simulated images. A variable and randomized camera magnification was used to simulate variations in the EPID-to-source distance and the associated image magnification effects. 

WL fields defined either using cones, collimator jaws or MLCs, are generally small compared to the size of the EPID detector. The out-of-field region does not provide any useful quantitative or qualitative data and does not serve any predictive value for the DL algorithm. Therefore, the range of EPID magnification values were optimized to ensure only the MLC field and a small, randomized margin, was visible in the synthesized image, eliminating the need for image cropping as a separate post-processing step.

The space between the light source and the virtual EPID camera was filled with the relevant linac components necessary to simulate a WL QA field. 

The central fifteen MLC leaves on each bank of a Varian (Palo Alto, California) Millennium MLC system were each modelled as a rectangular plane of width 5 mm. The ends of the leaves were bevelled with a magnitude which was controllable via the python script to simulate the leaf end effects typically seen in our WL fields. The MLC leaves were assigned a black diffuse material designed to block and absorb any optical rays between the source and EPID without any reflection or transmission. The length and depth of the leaf had no effect on the resultant WL images in the current work as the synthetic images were cropped to a small region around the WL MLC field and complete absorption of the photons by the MLCs was simulated using optical material properties. 

Collimator rotation was simulated by adjusting the rotation axis of each leaf to be at a point on the central axis of the simulated linac, at the height of the MLC bank. All leaves were then grouped to a control object so that rotating the control object also rotated all MLC leaves about the central axis of the linac. This rotation was also randomized by the python script to ensure WL fields of different collimator angles were represented in the training data set. 

Since only the central fifteen MLC leaves on each MLC bank were included in the simulation, a ‘block-out' geometry was placed laterally beyond the MLC geometry to ensure no optical rays from the light source could scatter around the MLC leaves and contribute to the final simulated WL image projected on the EPID. MLC transmission was not simulated in the current work, but this could potentially be simulated using a combination of Blender’s transmission material property or the principled volume (or volume scatter) material. 

Our custom QA phantom which is used for: kV/MV image scaling and quality checks, WL, “Snooker Cue” and light-radiation coincidence tests, was modelled in Blender in terms of geometric scale and BB placement. The QA phantom was simulated as a rectangular phantom of side length 12 cm and thickness of 2 cm. The phantom was assigned a transparent material with an index of refraction of 1.490 to match a Perspex material. X-ray phantom scatter was simulated by adjusting the roughness parameter of the transparent/glass material in Blender which had the effect of varying the penumbral width of the WL field. The six BBs in the QA phantom were simulated as spheres of nominal diameter 5 mm and a diffuse material with an optical transmission value which varied from 0.88 to 0.98. The diameter of the BBs was uniformly randomised by ± 20\% from the nominal diameter of 5 mm. These values were determined through experimentation to approximate the radiographic appearance of the BB shown in our WL images acquired through routine QA. While only the central BB can be seen in the simulated WL fields, the five other BBs were included in the modelled phantom for use in future work. The lateral position of the QA phantom relative to the EPID was randomised for each synthetic image with displacements of up to 7 mm from the isocentre. 

The couch and resultant couch transmission effects were not modelled directly in the current work as it was not essential to simulate the visual appearance of the WL fields.

\begin{figure}
	\centering
	\includegraphics[width=12cm]{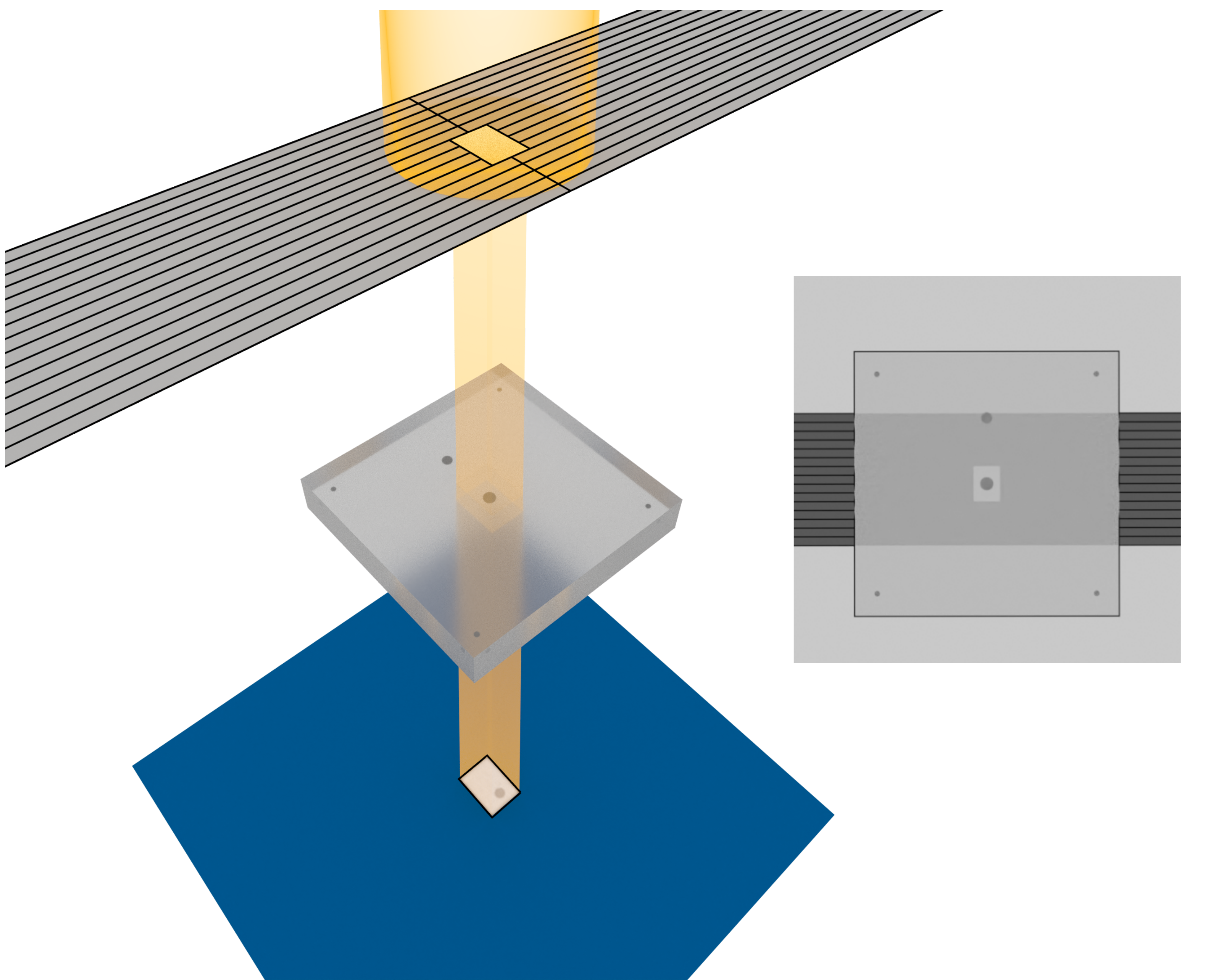}
	\caption{Geometric setup of Blender scene for synthetic image generation. Left: From top-bottom: MLC leaves, QA phantom (and 6 ball bearing targets), EPID panel (blue). The yellow region illustrates the path of the optical photons through the MLC leaves with an example of a synthetic WL image on the EPID (blue). Right and inset: Beams-eye-view of the QA phantom from the EPID showing collimation by the MLC leaves and ball bearing placements.}
	\label{fig:fig2}
\end{figure}

A rendering of the geometric setup of the Blender scene can be seen in Figure \ref{fig:fig2}. This figure does not show the ‘block-out' geometry beyond the lateral extent of the MLC leaves. Table \ref{tab:table1} lists all model parameters which were controlled and randomised by the python script in the current work.

\begin{table}
	\caption{A list of the parameters which were uniformly randomised by the custom Python script and the range of values for each in the current work. Note: for the x-ray source radius, the range of values do not (and should not necessarily) correspond with a typical Linac}
	\centering
	\begin{tabular}{lll}
		\toprule 
		Linac Parameter     & Minimum Value     & Maximum Value \\
		\midrule
		QA Phantom Centre Displacement from Isocentre & 0 mm  & 7 mm     \\
		EPID Lateral Displacement from Isocentre     & 0 mm & 5 mm      \\
		Ball Bearing Diameter     & 4 mm       & 6 mm  \\
		Ball Bearing Optical Transmission     & 88\%       & 98\%  \\
		MLC Leaf End Bevel     & 0 mm       & 1 mm  \\
		X-Ray source intensity     & 1 mW       & 10 mW  \\
		X-Ray source radius     & 5 mm       & 15 mm  \\
		QA Phantom Optical Roughness     & 5\%       & 7\%  \\
		EPID Orthographic Scale (zoom)     & 15      & 30  \\
		Collimator Rotation     & 0$^{\circ}$       & 180$^{\circ}$  \\
		\bottomrule
	\end{tabular}
	\label{tab:table1}
\end{table}

\begin{figure}
	\centering
	\includegraphics[width=8cm]{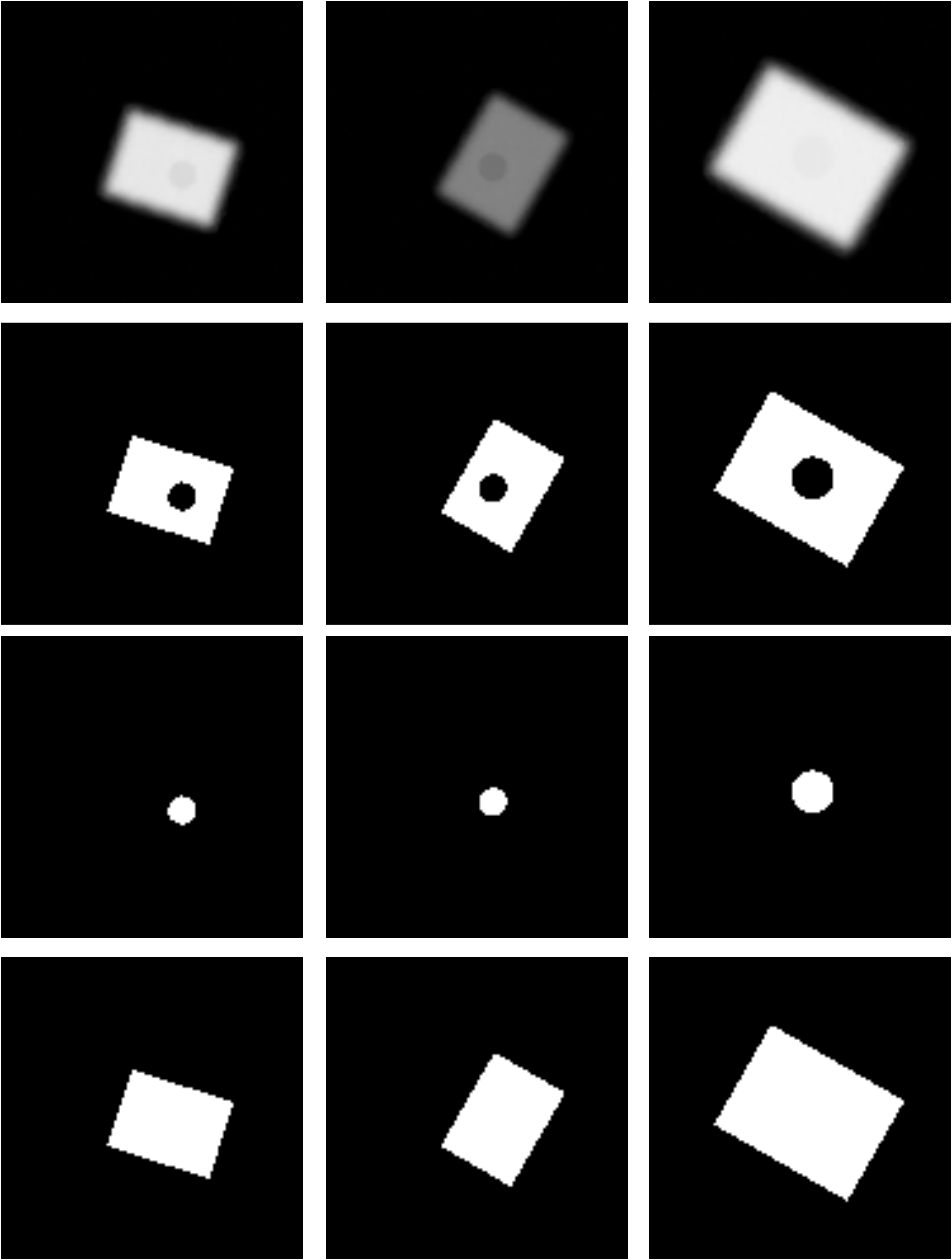}
	\caption{A random sample of three synthetic WL images and corresponding labelled masks. The top row shows three randomised synthetic WL images. Rows 2-4 show the corresponding pixel-wise labelled masks of the MLC field minus the ball bearing, ball bearing only and MLC Field Only.}
	\label{fig:fig3}
\end{figure}

Blender was configured using the compositor tool to generate pixel-wise masks corresponding to specific linac and phantom objects to produce matched input/output training data pairs for training of the DL model. The first output layer represented the synthetic MV WL image shown in the top row of Figure \ref{fig:fig3}). Layers two, three and four were configured to generate relevant “ground-truth” masks needed to train the DL model. The first mask layer was configured to generate a mask of the MLC field minus the BB (shown in the second row of Figure \ref{fig:fig3})). The second mask represented the position of the BB only and the final mask layer represented the position of the MLC field only (shown in the bottom row of Figure \ref{fig:fig3})). Layer 4 was used for testing and debug purposes only and was not an input to the DL model. 

The synthetic WL images and corresponding masks were rendered and exported as 8-bit grayscale images using the python script. Similar to typical Monte Carlo radiation transport simulations, Blender’s path-tracing render engine produces statistical noise in the output image which is proportional to the number of optical samples/histories. Blender includes several options to artificially remove statistical noise in the rendered image using Intel’s Open Image Denoise library \cite{RN46} or Nvidia OptiX Denoiser \cite{RN47} and thus reduce render times. In the current work, this option was disabled to allow some statistical noise to be present in each WL image thus improving consistency with a typical EPID image. Each WL image was rendered using 300 samples/histories (determined through trial-and-error approach) which resulted in a suitable balance between render time and statistical noise. 

The custom script was used to generate 1500 randomised synthetic WL images and corresponding masks each with dimensions of 120 $\times$ 120 pixels. These image sets were automatically named and exported to separate folders for training, validation, and testing of the DL model. Each of the simulated WL images and corresponding masks, were generated using a desktop PC (Intel Core i5-6500T) with no dedicated GPU, requiring on average 3-4 seconds to generate each image set. 

\subsection{Image Processing}

Prior to training the DL model, the synthetic WL images and corresponding masks were imported using a python script and the OpenCV toolkit \cite{RN45}. The mask corresponding to the MLC field without the BB was then combined with the BB only mask and assigned separate class labels to differentiate the BB from the MLC field. This was performed to ensure that each pixel in the combined label image corresponded to one class only, either MLC field, BB or neither. This WL image and combined mask were used as the input and output respectively during training of the DL model.

A total of 1500 synthetic images were imported, individually normalised to the maximum pixel intensity of each image, and then divided into training and validation data sets using a 75\%/25\% ratio. The mean pixel intensity of the training images was subtracted from each training and validation image and then divided by the standard deviation of the pixel intensities of the original training dataset. 

Data augmentation was implemented using Keras’ \cite{RN49} image generator in the form of horizontal and vertical flips, horizontal and vertical translations (±10\%) and zoom (±10\%) of the original WL training images and corresponding masks to artificially increase the training dataset size.

\subsection{Deep Learning Architecture and Training}

\begin{figure}
	\centering
	\includegraphics[width=12cm]{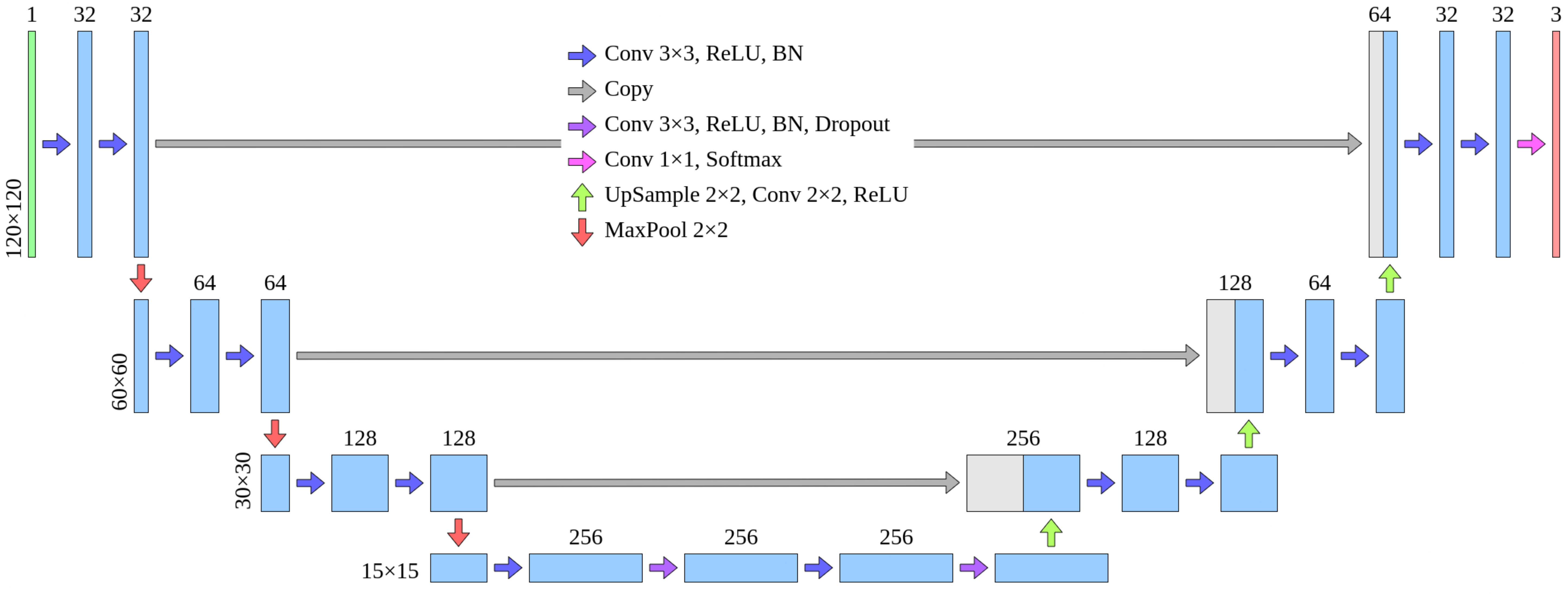}
	\caption{U-Net architecture used for DeepWL in the current work.}
	\label{fig:fig4}
\end{figure}

A three level U-Net  was implemented in the current work using Keras with a TensorFlow \cite{RN50} backend. Batch normalisation and dropout layers with a dropout rate of 0.5 were added to the network architecture to improve network stability and robustness. The U-Net architecture consisted of three up-sample and down-sample operations in the encoder and decoder respectively with four levels of feature resolution. The architecture of the DeepWL network is shown in Figure \ref{fig:fig4}). 

Prototyping of the initial DL model in terms of optimising network hyper-parameters, input image size and loss function type, was performed using Google Collaboratory \cite{RN51} and a GPU accelerated Jupyter notebook. Loss functions investigated included various combinations of categorical cross entropy, dice loss and focal loss. Several networks were trained with different input image sizes ranging from 80 $\times$ 80 to 120 $\times$ 120 pixels and hyper-parameters including constant and variable learning rates. 

The final network was trained using a combined categorical cross entropy and dice loss function, in a 1:2 weighting respectively, and the Adam optimiser function using a typical desktop PC using a CPU only. Previous studies have demonstrated that a dice loss function may reduce difficulties in training DL models on data with a large class imbalance \cite{RN66}. At each convolutional layer, 'same' padding was used as required, such that the layer's output size matches its input size. Network weights were initialised using HeNormal initialisation and a rectified linear unit (relu) activation was used at each layer except for the output which used a SoftMax activation. 

A variable learning rate was also implemented with a learning rate which reduced from 10$^{-4}$ to 10$^{-7}$ upon approaching the plateau of the training loss function. No transfer learning techniques were utilised in the current work. Model performance was evaluated during training in terms of pixel wise classification accuracy. 

\subsection{Model Validation on Synthetic Data}
Once the DeepWL model had been trained, one-hundred independent synthetic WL images were rendered using Blender and used to test the model’s performance. Python’s SciKit-Image module \cite{RN44} was used to find the un-weighted (binary) centre of mass location of the BB and MLC masks for the ground-truth and predicted masks , enabling the magnitude of the displacement of the BB from the MLC field to be calculated and analysed.

The Dice similarity coefficient (also known as the Sørensen–Dice coefficient or DSC) was also evaluated for each of the predicted BB and MLC field masks using the Blender predicted masks as the ground-truth masks, for comparison. The DSC is a metric used to assess similarity between two data sets and is defined as twice the number of pixels (or voxels) common to both label maps divided by the sum of the number of pixels in each label map.

The BB to MLC field displacements for the ground truth masks and DeepWL predictions were calculated and compared using a statistical t-test and Pearson’s correlation coefficient. 

\subsection{Evaluation on Measured Winston Lutz Data}

Once the model was trained and tested on synthetic WL data, experiments were performed to determine whether the model was sufficiently general to accurately segment the MLC field and BB of WL fields measured on a real EPID panel.

WL images in DICOM format, measured as part of routine linac QA were retrospectively collected for five Linacs in our department for the year 2020. Two of the linacs were Varian Truebeam Stx (AS1200 EPID panel), two were Varian Clinac iX (one with AS1000 and one with AS500 EPID panel) and the fifth was a Varian Trilogy (AS1000 EPID panel). 

A total of 228 EPID images were collected for the 5 linacs, each with varying gantry, collimator, and couch rotations. Of the 228 images, 43 were from AS1200 EPID panels, 109 were from AS1000 panels and 76 were from the AS500 panel. Images from the AS1200, AS1000 and AS500 panel had dimensions of 1190 $\times$ 1190 pixels, 1024 $\times$ 768 and 512 $\times$ 384 pixels, respectively. Each WL image was acquired with the EPID panel positioned at 50 cm posterior to the mechanical isocentre of the linac in accordance with our QA protocol. 

In addition to collecting routinely measured WL images, several WL images were acquired on our Truebeam linac with intentionally large BB offsets ranging from 1.5 – 2.0 mm (defined at the isocentre) to evaluate the robustness of the CED and DeepWL methods to outliers. 

These WL EPID images were then imported in DICOM format using the Python “pydicom” module. Our standard code for analysing WL images in our department begins by cropping the images to a region which extends just beyond the MLC field (Figure \ref{fig:fig1})). Due to the differences in image size and resolution of each of the three EPID panels, the region that is cropped is specific to the EPID imager that was used to acquire the image. The images from the AS1200, AS1000 and AS500 EPID panels were cropped to a region of 120$\times$120 pixels, 110$\times$110 pixels and 70$\times$70 pixels respectively and centred on the central pixel of the original image. These same cropped regions were used for both the CED and DeepWL methods. 

Since the DeepWL model was trained on 120$\times$120 pixel images, the images for the AS1000 and AS500 EPID panels were subsequently resampled to 120 $\times$ 120 pixels using OpenCV’s  inter-area interpolation. The CED method was evaluated on the cropped images while DeepWL was evaluated on the up-sampled images to compare our standard QA practice with the proposed DeepWL method. 

The CED method calculates the BB and MLC field centre using the following method. First, the CED algorithm from the SciKit-Image module is evaluated on the cropped WL image. Label maps are then generated from the predicted edges providing they meet the required “connectivity” requirements, that is, pixels representing edges of the BB or MLC are connected to neighbouring pixels. 

A processing step is then applied to remove predicted labels which are too small to be either the BB or MLC label. A final post-processing step is then applied which filters predicted labels by eccentricity (how elliptical or circular the label is) which assists in removing unwanted labels and preserving only the BB and MLC labels. Finally, the un-weighted centre of mass location of the BB and MLC labels are calculated from their respective label maps and their relative displacement in millimetres (at the linac isocentre) is calculated by scaling the predicted displacement measured in pixels by the EPID pixel size (mm/pixel) and a magnification factor which accounts for the EPID distance from isocentre. 

The same WL images were then evaluated using DeepWL using the following method. The cropped and up-sampled WL images were normalised to the maximum pixel intensity. Images taken on the AS1000 and AS500 EPID panels were inverted and then re-normalised to ensure consistency in the appearance of all EPID images. The mean of the normalised synthetic training data was subtracted from each normalised image and then divided by the standard deviation of the training data. The trained DeepWL model was then loaded with associated weights and each of the 228 images were evaluated by the model producing a probability map with dimensions of 120$\times$120$\times$3. The third dimension represents the three possible class probability maps predicted by DeepWL: out-of-field, radiation field minus BB and BB. A pixel map of dimensions 120$\times$120$\times$1 was then generated by assigning a class to each pixel in the WL image based on the class with the highest probability for a given pixel. 

The pixel corresponding to the centre of the BB and MLC field was then calculated by finding the un-weighted centre of mass of the BB label map and the centre of mass of the inverse of the out-of-field map respectively, since the radiation field class excludes the BB. The displacement of the BB and MLC field centres (measured in pixels) was then scaled by the EPID pixel resolution, EPID magnification factor and the ratio of the cropped image dimensions to 120 pixels to compensate for the image up-scaling. 

The BB to MLC displacements, predicted by the CED and DeepWL methods, were compared for each image. A two-sample t-test was used to assess whether the mean of the predicted offsets of all images using the two methods were statistically different. 

The BB centre predictions of the CED and DeepWL methods were subsequently compared to manually annotated ground truth values to resolve any discrepancies in their predictions. Manual annotations were performed for each of the 228 cropped WL images using ImageJ’s multi-point tool and manually windowing each image in the stack for optimal viewing of the BB. Manual annotations were performed independently of the CED or DeepWL predictions and compared with the other two methods. Agreement between the different methods of BB localisation were analysed using Pearson’s correlation coefficient and statistical t-tests. 

We hypothesised that the DeepWL model may also produce more accurate segmentations of the BB having “learnt” that the BB is always circular, in addition to considering the relative pixel intensities of the image. Unlike the CED method which relies entirely on the relative pixel intensities of the BB compared to the MLC field. 

To test this hypothesis, the morphology tools of Python’s SciKit-Image module were used to measure the eccentricity of the predicted BB contours. This was used to establish whether DeepWL produces BB labels more consistent with the shape of the actual BB in the phantom compared with CED. Eccentricity of 0 implies a perfect circle, Eccentricity > 0 implies a more elliptical contour. Statistical analysis was performed to compare the mean eccentricity for all images predicted using the CED vs DeepWL method. 

\section{Results}

\subsection{Training and Model Validation}

\begin{figure}
	\centering
	\includegraphics[width=8cm]{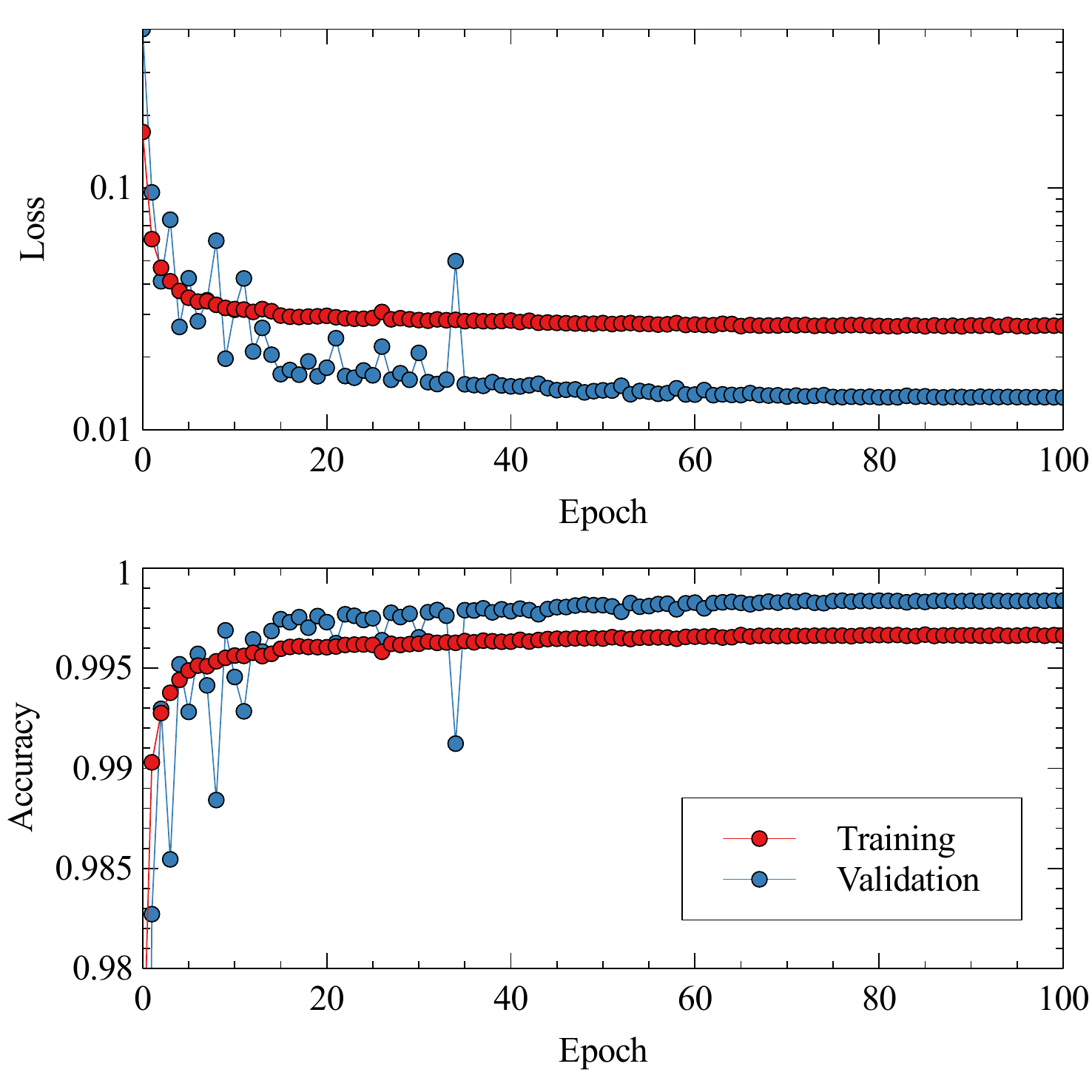}
	\caption{Training/Validation Loss and Accuracy for the final DeepWL trained model.}
	\label{fig:fig5}
\end{figure}

The final model was trained using a CPU rather than GPU, requiring approximately 15 hours to train for 180 epochs, with the model converging after approximately 50 epochs. The training/validation loss and training/validation accuracy for the model is shown in Figure \ref{fig:fig5}). This result demonstrates this approach is feasible for most radiation oncology departments without dedicated GPU hardware. For comparison, training of the model during the prototyping stage on Google Collaboratory using a GPU required approximately 12-15 minutes to complete.

\begin{figure}
	\centering
	\includegraphics[width=8cm]{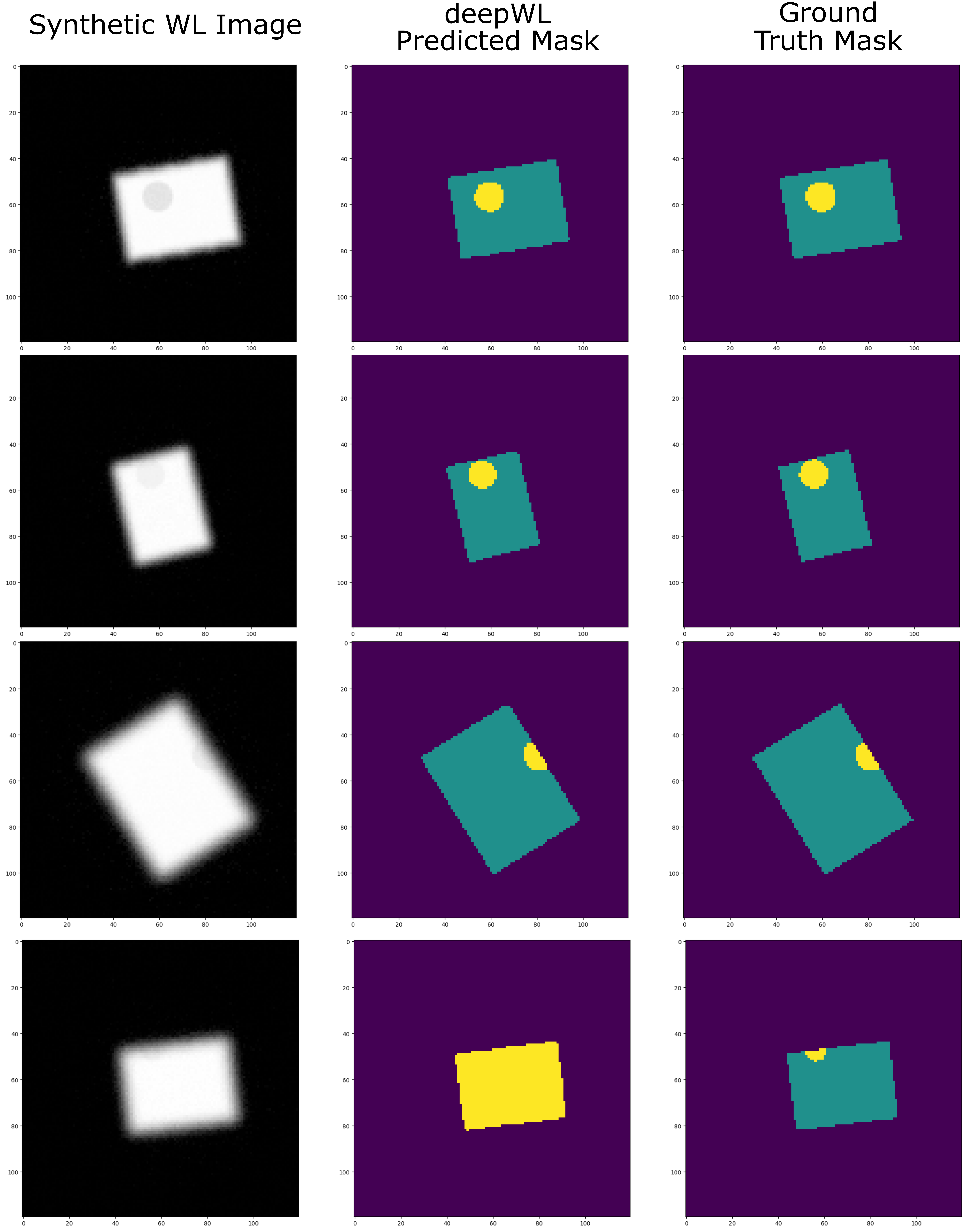}
	\caption{Three examples of synthetic WL images with the deep WL predicted masks and ground-truth masks. The fourth row shows the only example where DeepWL failed to segment the ball bearing or MLC field in the testing data}
	\label{fig:fig6}
\end{figure}

The model was then evaluated on 100 independent testing WL images. Four of these images are shown in  Figure \ref{fig:fig6}). The mean DSC for the MLC field and BB were 0.994 ± 0.002 and 0.964 ± 0.09 respectively indicating a high degree of similarity between the ground truth and predicted masks. However, DeepWL was unsuccessful in predicting the masks correctly for one of the hundred test images due to insufficient contrast between the MLC field and BB. This does not indicate a limitation of the model, but rather the BB transmission property in the Blender scene was too high for this synthetic image. This is demonstrated in row 4 of Figure \ref{fig:fig6}).

\begin{figure}
	\centering
	\includegraphics[width=8cm]{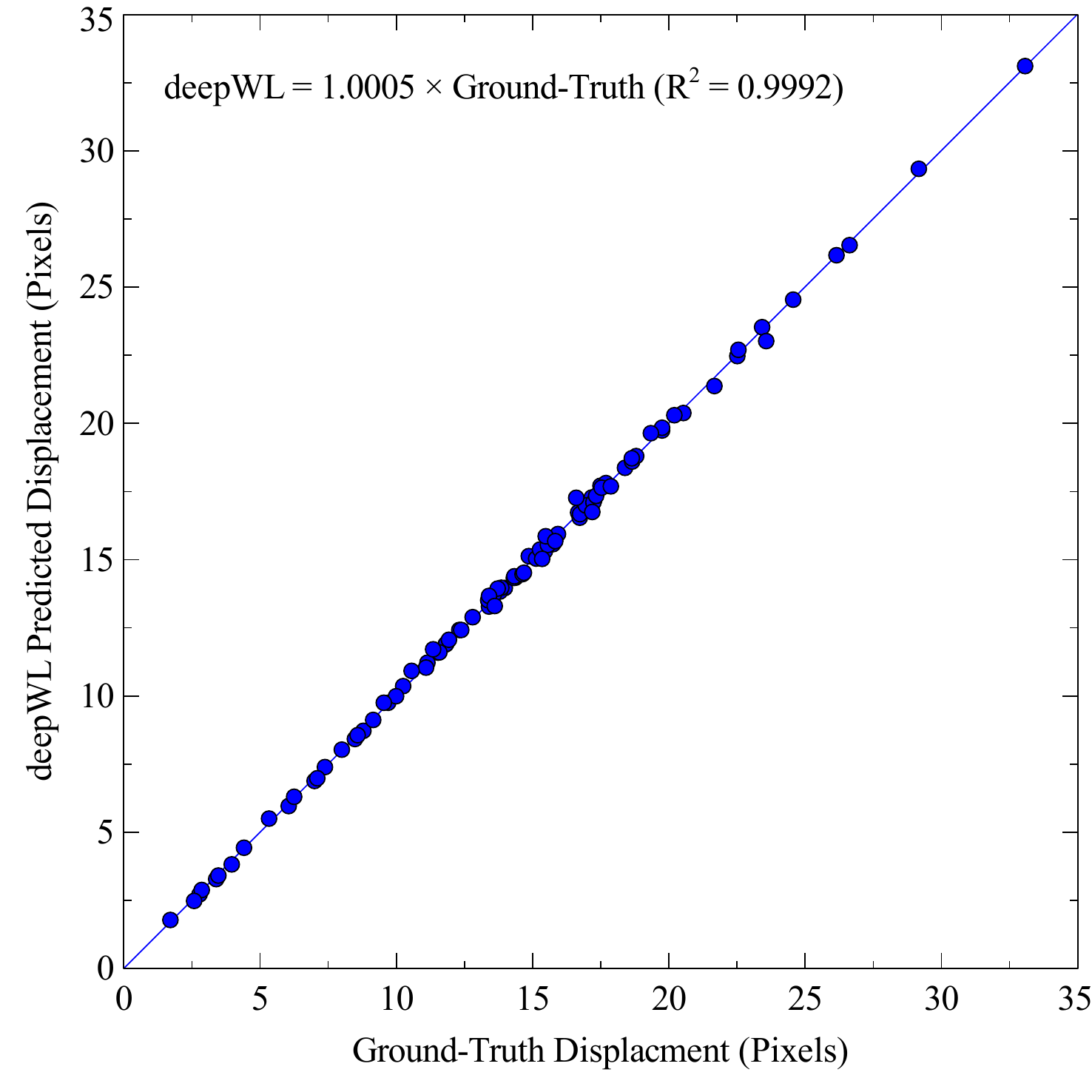}
	\caption{Ground truth versus DeepWL predicted ball bearing to MLC field displacement for 99 independent test images}
	\label{fig:fig7}
\end{figure}

The predicted BB to MLC field displacements from the DeepWL model were compared with the ground truth displacements and plotted in Figure \ref{fig:fig7}) for the remaining 99 test images. A linear regression was performed between these displacements with the intercept set to zero (i.e. a centred BB and MLC field should result in zero displacement in both the ground truth and predicted cases). The slope of this regression was calculated to be 1.0005 with a R$^{2}$ coefficient of 0.999 with a Pearson’s correlation coefficient of 0.999 indicating a high degree of agreement between the DeepWL predicted displacement and ground truth values.

A two-sample, two-sided, t-test assuming unequal variances was performed between the mean ground truth displacement (mean = 14.27 pixels) and mean DeepWL displacement (mean = 14.28 pixels). The hypothesised mean difference in the predicted displacements was zero. The p-value was calculated to be 0.99 at a 95\% confidence indicating that there was no statistically significant difference in the mean predicted displacements between the ground truth and DeepWL contours.

This data provided evidence that the DeepWL model was trained successfully and was capable of accurately predicting the BB and MLC masks on previously unseen synthetic WL data exported from Blender. 

\subsection{Evaluation on Measured WL Data}

\begin{figure}
	\centering
	\includegraphics[width=8cm]{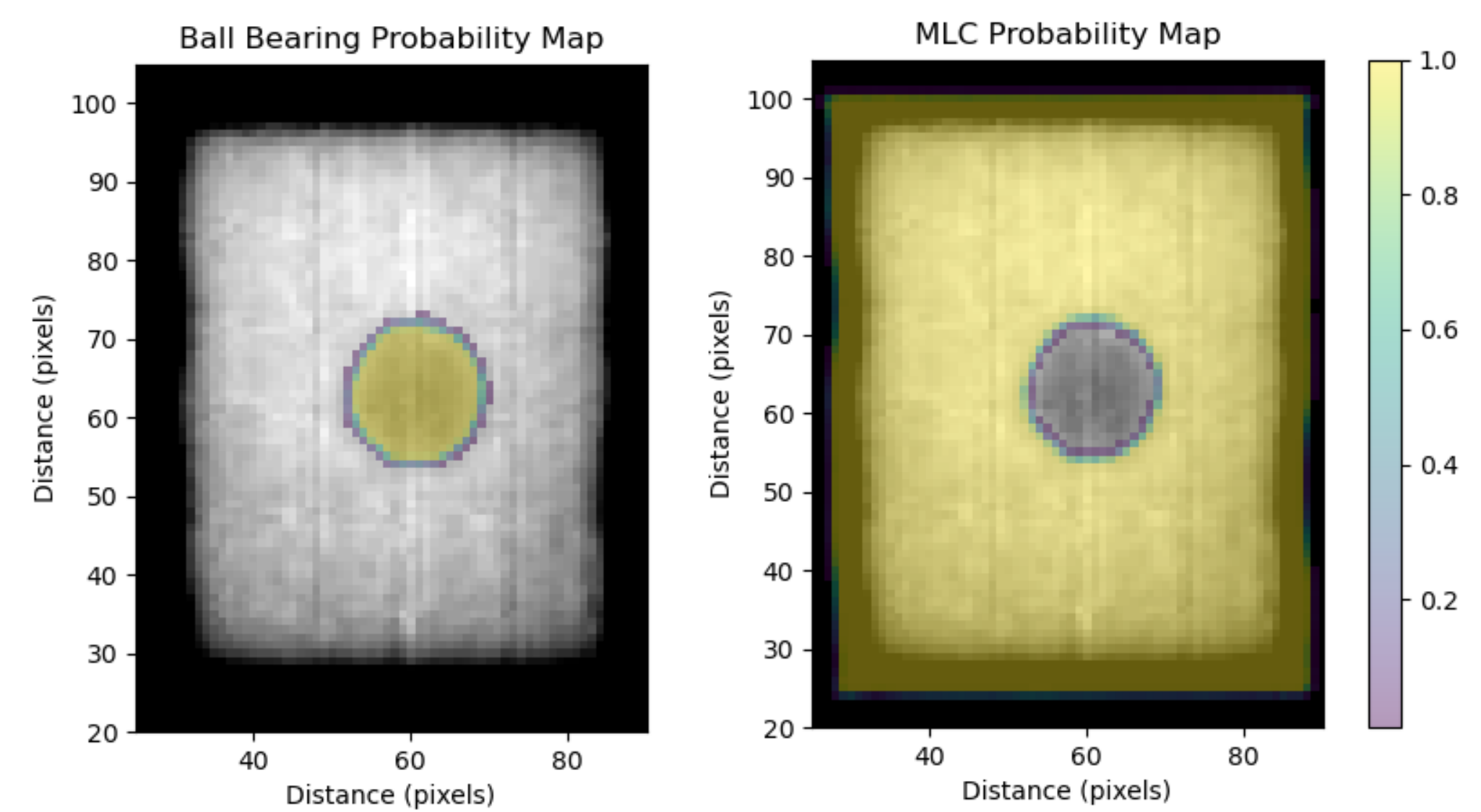}
	\caption{The DeepWL predicted probability map for the ball bearing and MLC field for a sample WL image}
	\label{fig:fig8}
\end{figure}

228 measured WL images from 5 Linacs with 3 different EPID panels and 2 MLC designs were evaluated using the model. The probability heatmaps for the BB and MLC field for one of these images is shown in Figure \ref{fig:fig8}). The CED method required on average 13.8 ± 2.8 µs to segment and calculate the displacement for each image. The DeepWL model required on average 4.28 ± 0.80 µs to calculate the displacement of each image once the model had been loaded and the label maps generated for all 228 images. The time required to generate the label maps for all 228 images simultaneously was 14.1 seconds with a single image requiring 0.97 seconds to evaluate. In addition, the time required to load the DeepWL model and associated weights was 1.1 seconds. 

\begin{figure}
	\centering
	\includegraphics[width=8cm]{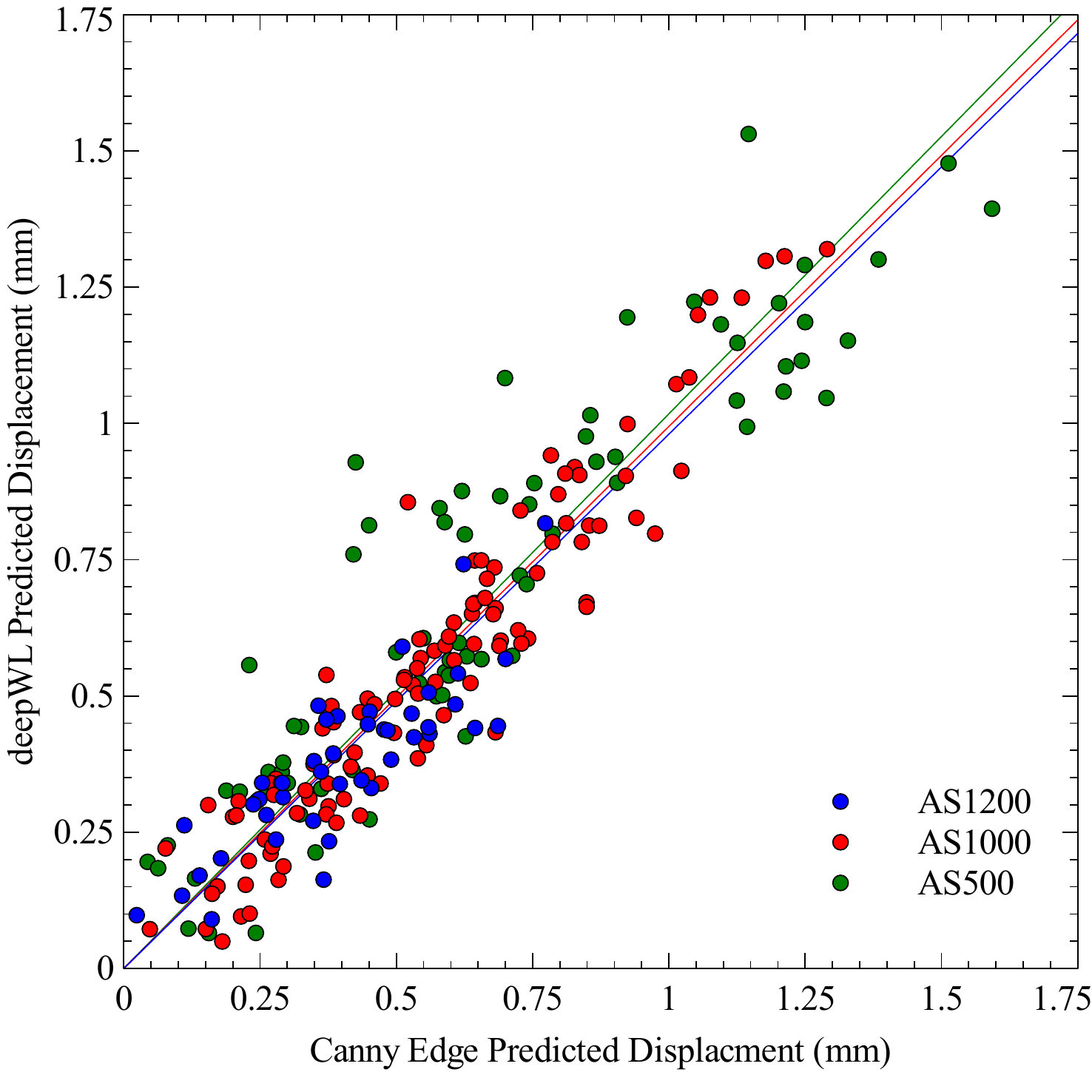}
	\caption{CED predicted BB-MLC displacements vs DeepWL predicted displacement (mm) grouped by EPID panel (AS1200, AS1000, AS500)}
	\label{fig:fig9}
\end{figure}

The BB to MLC field displacement predicted by the CED and DeepWL model for each EPID type is plotted in Figure \ref{fig:fig9}). Each EPID panel showed a 1:1 linear relationship between the two analysis methods. Visually, there appears to be better agreement for the AS1200 and AS100 panels in terms of smaller residuals from the line of unity and the lower regression coefficient for the AS500 panel (R$^{2}$ = 0.81 for AS500 versus R$^{2}$ = 0.88 and R$^{2}$ = 0.89 for the AS1200 and AS1000 respectively).

\begin{figure}
	\centering
	\includegraphics[width=8cm]{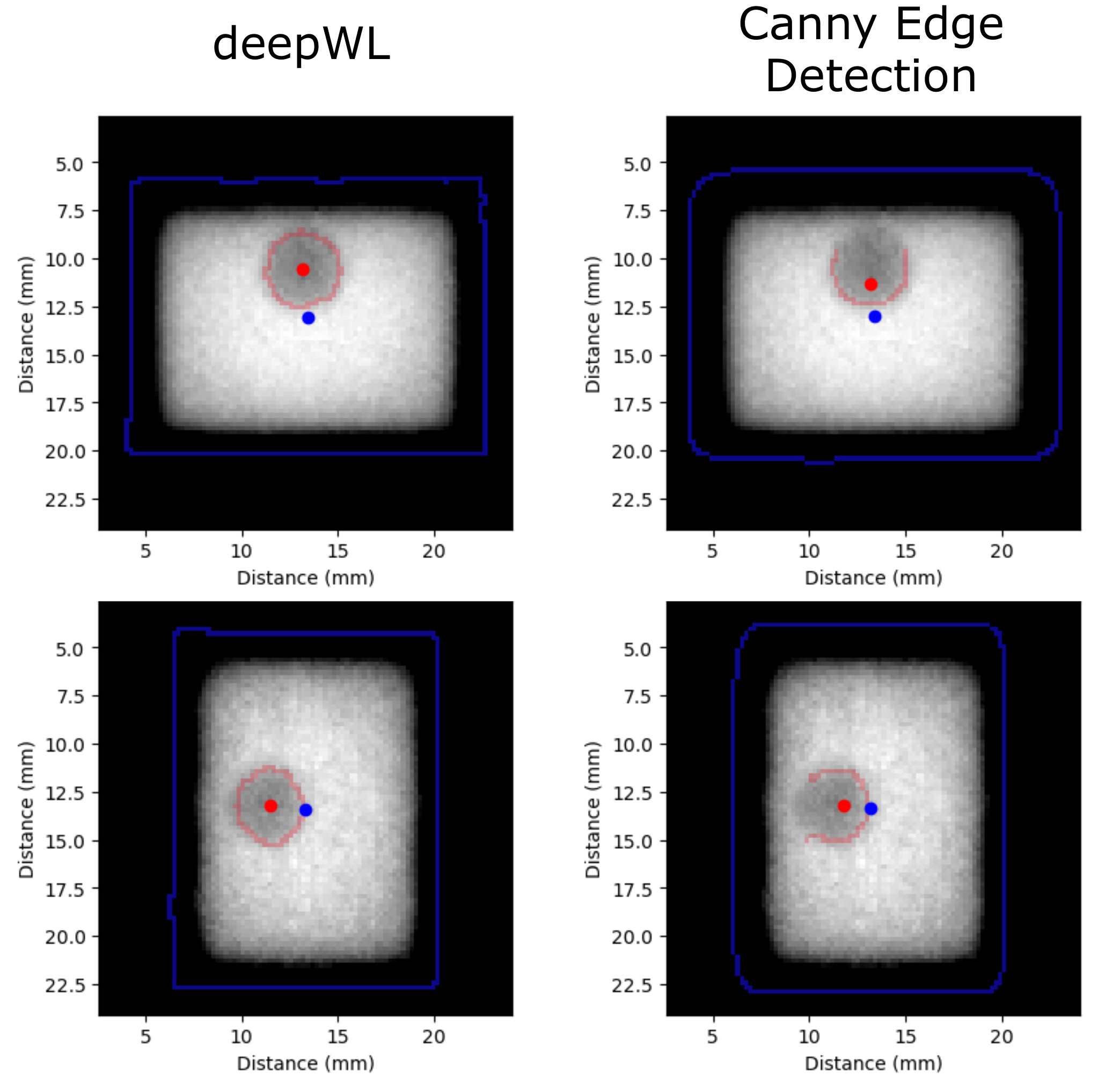}
	\caption{Examples of WL images with intentionally large offsets to demonstrate robustness of DeepWL model compared with CED. Images have been windowed for optimal BB to MLC field contrast.}
	\label{fig:fig10}
\end{figure}

Several tests were conducted using intentionally large, BB offsets in randomised directions of up to 2 mm from isocentre, to test the robustness of each method to outliers where the BB approaches the MLC field edge, and the BB becomes increasingly difficult to separate from the field edge. Some examples from this test can be seen in Figure \ref{fig:fig10}). The CED approach failed to correctly segment the BB as it approached the edge of MLC field resulting in an underestimation of the true BB offset relative to the MLC field centre. The DeepWL method correctly segmented the BB in each case, maintaining the correct BB shape and correctly predicting the true offset of the BB. This example demonstrates that the DeepWL method may be more robust to such outliers whilst the CED may produce spurious predictions under these extreme circumstances. 

A two-tail t-test was performed on the mean displacements predicted by the two models for all 228 routine QA images assuming un-equal variances to determine whether there was a statistically significant difference in the predictions of the two models. The mean of the CED method (mean = 0.566 mm) and the mean of the DeepWL method (mean = 0.571 mm) was determined to be non-significant at a 95

\begin{figure}
	\centering
	\includegraphics[width=8cm]{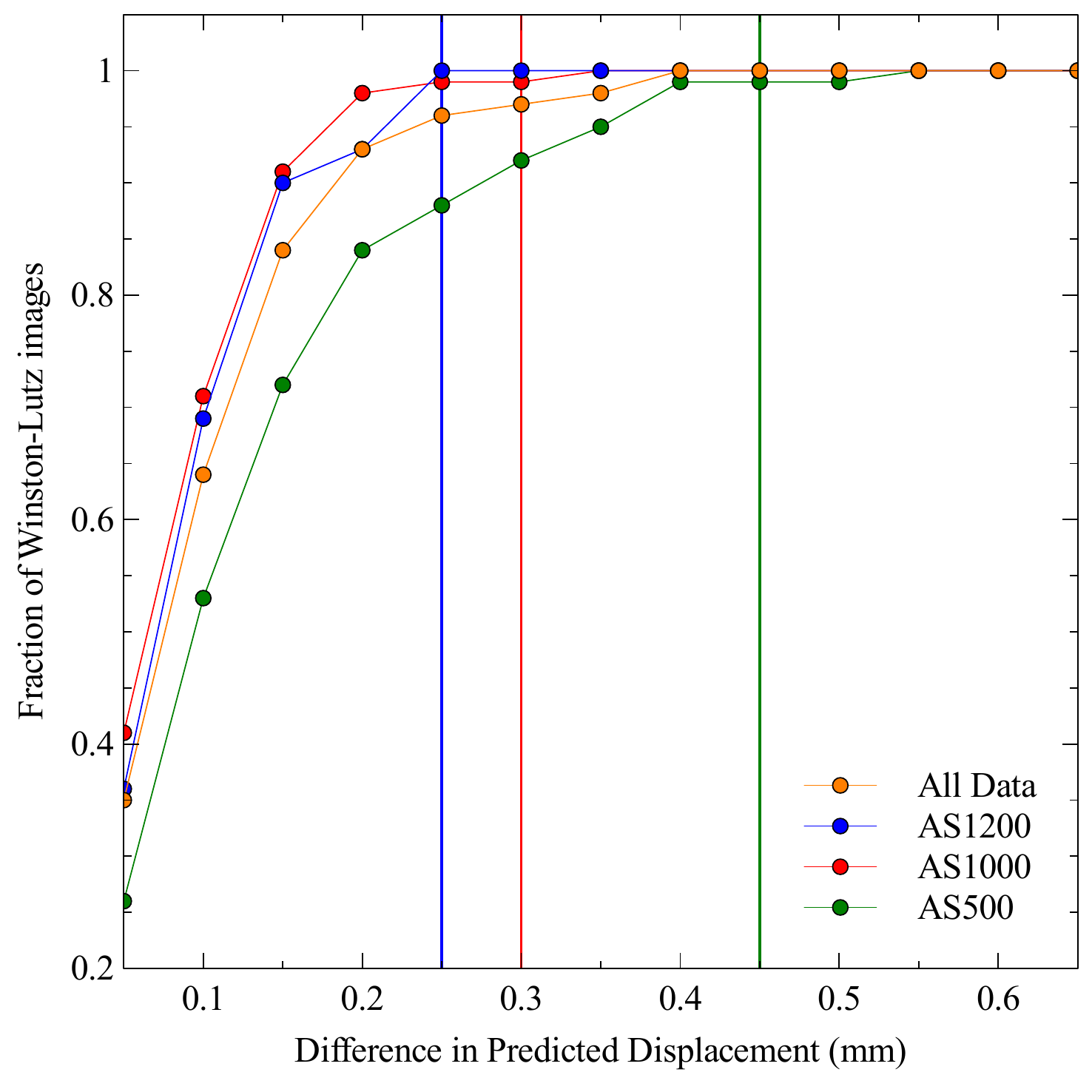}
	\caption{Examples of WL images with intentionally large offsets to demonstrate robustness of DeepWL model compared with CED. Images have been windowed for optimal BB to MLC field contrast.}
	\label{fig:fig11}
\end{figure}

To further investigate the significance of the difference in the predicted displacements, the pair-wise difference in predicted displacements for each of the 228 images was calculated and a histogram was produced with a bin size of 0.05 mm. A cumulative distribution is shown in Figure \ref{fig:fig11}). The fraction of images whose predicted displacements agreed to within the spatial resolution of the corresponding EPID at the isocentre was calculated and tabulated in Table \ref{tab:table2}. The spatial resolution of the AS1200, AS1000 and AS500 EPID panels at the isocentre are 0.22 mm, 0.26 mm, and 0.43 mm respectively and these values were assumed to be the smallest difference in predicted displacement deemed significant for this analysis. When all 228 images were analysed together, all predicted displacements between CED and DeepWL agreed within 0.55 mm.

When the images were grouped by EPID panel and analysed separately, it was found that the CED and DeepWL predictions for images on the AS1200 panel agreed in all cases to within the spatial resolution of the EPID. The results for the AS1000 panel showed 99.08\% agreement to within its corresponding spatial resolution. However, when the BB and MLC field displacements were analysed separately, 100\% agreement was achieved within the spatial resolution of the EPID. The AS500 panel showed reduced agreement of 98.65\% at the corresponding resolution of 0.43 mm. The predictions were again separated into BB and MLC field displacements revealing that the reduced agreement was due to differences in BB displacement predictions.

\begin{table}
	\caption{Summary of DTA analysis. Percentage of WL images within spatial resolution of each EPID panel, BB to MLC displacement, BB displacement and MLC displacement from central pixel}
	\centering
	\begin{tabular}{llll}
		\toprule 
		 & AS1200     & AS1000     & AS500 \\
		\midrule
		\% Agreement of BB- MLC Displacement & 100\%  & 99.08\% & 98.65\%    \\
		\% Agreement of BB Displacement & 100\%  & 100\% & 98.65\%    \\
		\% Agreement of MLC Displacement & 100\%  & 100\% & 100\%    \\
		\bottomrule
	\end{tabular}
	\label{tab:table2}
\end{table}

To resolve which of the two methods (CED or DeepWL) were predicting the BB location with higher accuracy, manual annotations of the BB location for all 228 WL images were performed by a medical physicist using ImageJ’s multi-point tool. The Pearson’s correlation coefficient was calculated between the DeepWL, CED and manual BB displacements (measured from the central pixel of the image). A consistently higher Pearson’s coefficient was found on each EPID panel for the manual versus DeepWL predicted BB displacements compared with the manual versus CED predictions. Whilst the difference in correlation between these groups appears insignificant for the AS1200 and AS1000 EPID, the correlation coefficient was higher for the DeepWL versus manual compared to the CED versus manual for the AS500 panel with values of 0.88 and 0.79, respectively. These results have been tabulated in Table \ref{tab:table3}.

\begin{figure}
	\centering
	\includegraphics[width=8cm]{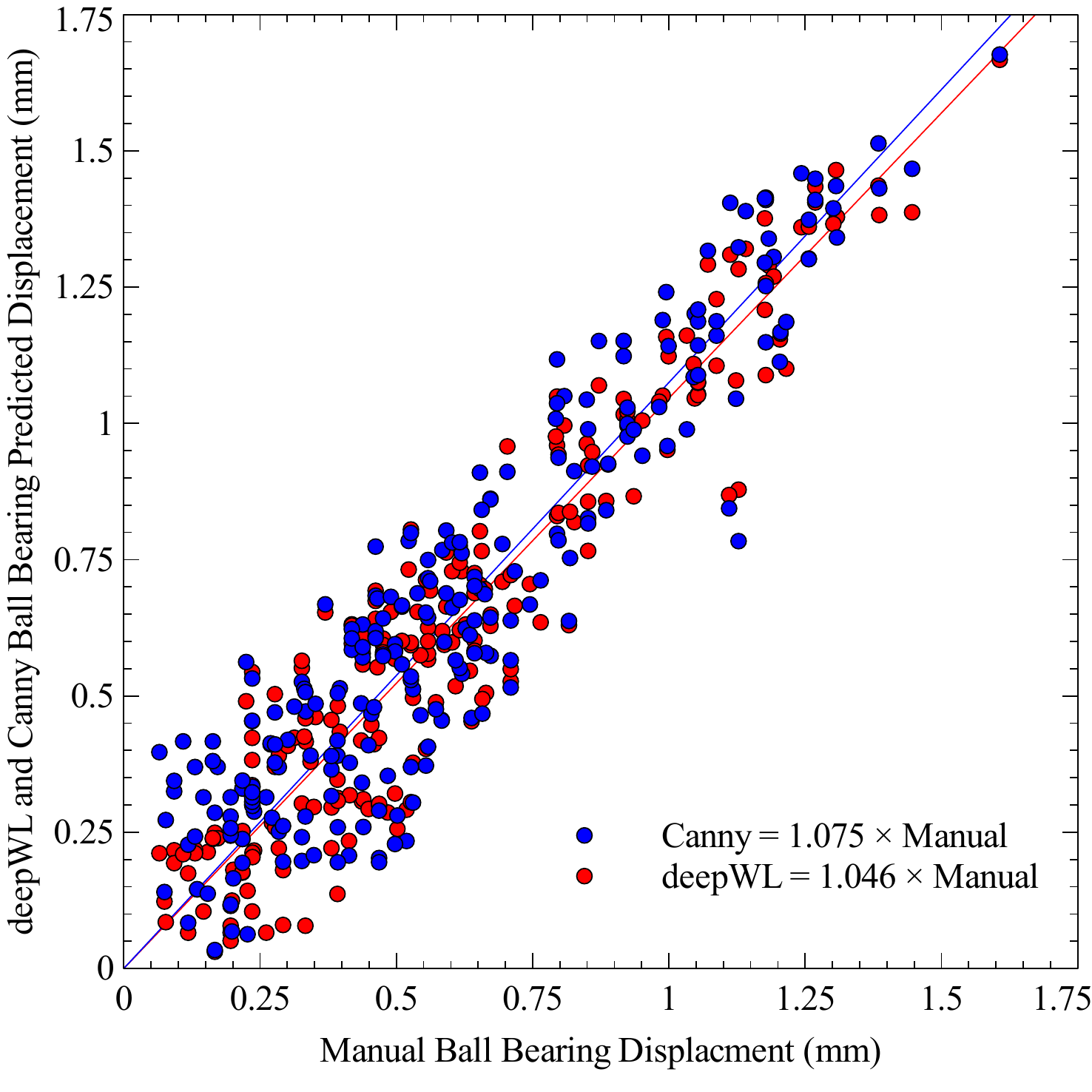}
	\caption{Manually annotated displacement of ball bearing from image centre (mm) vs DeepWL and CED predicted displacement (mm) for all WL images. The intercept has been forced to be zero.}
	\label{fig:fig12}
\end{figure}

\begin{table}
	\caption{Pearson’s Correlation Coefficient: ball bearing displacement relative to central pixel.}
	\centering
	\begin{tabular}{lllll}
		\toprule 
		 & AS1200     & AS1000     & AS500  & All\\
		\midrule
		Manual vs CED & 0.90 & 0.95 & 0.79 & 0.92     \\
		Manual vs DeepWL & 0.94 & 0.96 & 0.88 & 0.94\\
		DeepWL vs CED & 0.98 & 0.98 & 0.89 & 0.98    \\
		\bottomrule
	\end{tabular}
	\label{tab:table3}
\end{table}

A statistical analysis was performed to determine if the mean BB displacements predicted by the CED, DeepWL methods and the manual annotations were significantly different at 95\% confidence for all 228 images. The analysis revealed no statistically significant difference in the mean CED predicted BB displacements compared to manual displacements (p=0.069 two-tail t-test) and no significant difference for the DeepWL vs manual predictions (p=0.34 two-tail t-test).

A statistical analysis of the regression of each of the predicted displacements was also performed to assess the correlation. The coefficients for the linear regression between the manually annotated BB displacements and CED predicted displacements were between 0.94 and 1.05 for the gradient with an offset of between 0.02 and 0.10 at a 95\% confidence. For the manual vs DeepWL predictions the gradient was calculated to be between 0.98 and 1.08 with an offset between -0.02 and 0.05 at a 95\% confidence. These results demonstrate that both CED and DeepWL correlate well with manual annotations, with evidence for higher predictive accuracy with DeepWL.

The eccentricity values for all 228 BB segmentations were calculated using SciKit-Image's eccentricity function. The mean eccentricities for BB segmentations predicted by the CED and DeepWL methods were 0.31 and 0.27 respectively. A one-tailed t-test was performed to determine if the mean eccentricity of CED was higher than DeepWL at a 95\% confidence. The difference in mean eccentricities was determined to be statistically significant with a p-value of 0.00025. A statistical analysis for the mean eccentricities calculated for each EPID panel separately also showed a statistically significant difference at a 95\% confidence.

These results suggest that DeepWL produces labels of the BB which are statistically more circular in shape and thus more consistent with the true shape of the target compared with the CED method. 

To determine if this difference in eccentricity of BB labels was a useful predictive measure of difference in BB displacement between the two predictive models, a statistical regression analysis was performed assuming an intercept of zero. The regression between the absolute difference in eccentricity and the absolute difference in predicted BB displacements between CED and DeepWL was performed at a 95\% confidence. The F-value for the regression was 235 with a corresponding probability of $7\times10^{-37}$ and therefore we reject the hypothesis that there is no relationship between these two variables. The t-statistic for the slope of this linear regression (gradient = (1.12, 1.46) 95\% confidence interval) was 15.3 with a p-value of p=$6\times10^{-37}$ indicating that absolute difference in eccentricity is a useful predictor of absolute difference in BB predicted displacement at a 95\% confidence. However, a subsequent statistical analysis of the correlation between BB eccentricities and the difference in the BB displacements (CED and DeepWL data grouped) compared with the manual BB annotations showed no statistically significant correlation at a 95\% confidence, with a p-value of 0.07 and 0.34 for the intercept and slope of the linear regression respectively. 

\section{Discussion}

In the current work, we were able to successfully demonstrate that a U-Net based DL model trained on synthetic EPID WL data, derived from an optical ray-traced simulation generalised to measured WL images from an EPID and was able to accurately segment the BB and MLC field. The predicted offsets were consistent with both manually annotated BB positions as well as CED-based predictions. 

We found that in general the CED and DeepWL methods of analysing EPID WL images agreed and that there was no statistically significant difference in the mean displacements between the centre of the MLC field and BB between predicted by each method at a 95\% confidence. This was further demonstrated from the strong degree of linear correlation between their predictions (R$^{2}$ =99.8) for the AS1200 and AS1000 panels with zero intercept and a >98\% agreement for all predictions within the spatial resolution of the EPID panels at the isocentre. 

However, further analysis revealed a larger discrepancy in the relative predictions on the AS500 EPID panel compared with the AS1200 and AS1000 EPID panels. This was likely caused by the reduced spatial resolution of the panel compared with the AS1200 and AS1000 models. A statistical comparison of both automated methods demonstrated that the DeepWL method produced predictions of the BB displacements which correlated better with manual annotations. This was further emphasised by a higher Pearson’s correlation coefficients for DeepWL versus manual for each EPID panel individually and as a group compared with CED. 

One reason for this improved performance was hypothesised to be due to improved BB contour circularity (reduced eccentricity). Statistical analysis revealed that DeepWL produced labelled masks of the BB region which were, on average, more circular than those predicted by the CED method and therefore, more consistent with the actual shape of the target. Whilst difference in eccentricity was shown to be a statistically useful variable in predicting differences in BB displacements between the two models, eccentricity was not useful in estimating differences from the manual annotations. 

Several of the analyses in this study assumed that the manual annotations of the BB locations could be considered accurate representations of the ‘ground-truth. However, the single pixel localisation method used to localise the BB centre manually was likely to have a high degree of uncertainty and is a limitation of this study. This uncertainty may have masked any observed statistical correlation with the CED and DeepWL predictions and makes it difficult to assess which automated analysis method produces more accurate predictions. 

Another, more precise, albeit time consuming method of manually annotating the BB would have been to manually select all pixels corresponding to the BB target and calculate the geometric mean of this region. Deciding which pixels correspond to this region on a low contrast image is also highly subjective, particularly near the edge of the BB and therefore has uncertainties as well. 

Another possible contributor to the reduced agreement in the predicted displacements for the AS500 EPID panel may have been the interpolation method used to re-sample the cropped image to the input dimensions of DeepWL. Images from the AS1000 and AS500 EPIDs were resampled from 110$\times$110 pixels and 70$\times$70 pixels respectively to 120 $\times$ 120 pixels as required by the DeepWL model. The larger degree of scaling required for the AS500 panel compared with the AS1000 may have contributed to the observed reduced performance on this EPID. Further investigation should be performed to separate the relative contribution of the image resolution from the interpolation method used on predictive accuracy. 

DeepWL was demonstrated to be more robust in scenarios where the BB displacement was comparable to the width of the MLC field which, in the case of the CED method, produces spurious and unreliable predictions. Whilst these extreme examples are outside the range of displacements typically seen as part of routine QA, it provides further evidence that the DeepWL model appears to consider the learnt shape of the BB target in addition to relative pixel intensities, resulting in improved predictive robustness to outliers. Similar observations have been made when comparing DL organ segmentation versus atlas-based segmentation in radiation oncology \cite{RN64}.

Early in the development of the model, attempts were made to locate the BB centre pixel using the pixel corresponding to the maximum predicted probability rather than the geometric centre of mass of the BB contour. Whilst this point always occurred in the BB region of the image, it did not correlate well with the true geometric centre of the BB. As a result, the geometric mean of the predicted BB mask was used to locate the BB for the remainder of the study, however this approach could potentially be useful with further investigation and development.

The DeepWL model was also evaluated on WL images acquired using a Brainlab (Munich, Bavaria) SRS WL phantom to test the generalisability of the model to other phantoms. However, neither DeepWL nor the CED method produced accurate predictions on this phantom. The CED method required a change in input parameters to correctly segment this phantom while the DeepWL code would require additional training on synthetic images of this phantom to give simultaneously accurate predictions on both phantoms. However, this raises an interesting possibility, that provided the model were trained on data which included the current variation of input parameters with the addition of multiple phantoms, the model may generalise sufficiently to accurately analyse WL images using a wide range of phantoms without a change in parameters. 

The training data generation technique utilised in the current work allows for the rapid development of synthetic data without requiring the time, expertise or geometric and compositional accuracy that would be required of a typical Monte Carlo radiation transport simulation. As was discussed in the first section of this article, our ‘in-house’ QA phantom was modelled in its entirety with the vision that one or more DL models could be trained to automate many of our routine linac quality assurance tasks. One key advantage of this approach is the improved accuracy and reduced time to produce high quality ground truth label masks for training DL models in radiation oncology. Manually annotating real WL images is time consuming and suffers significant uncertainties including inter observer variability.

The WL images analysed in the current work represent a small, cropped region of the EPID panel, in this case 120 $\times$ 120 pixels, which allows the DeepWL model to be trained quickly on non-dedicated hardware making this approach feasible for most medical physics departments without requiring specialised and often cost-prohibitive GPU hardware.

One limitation of the DeepWL model is the slower evaluation time per image, resulting from the time required to load the model weights and evaluate the model on each image. DeepWL has a small but non-significant advantage in evaluation time when performed on all 228 images as a group but analysis on a large group of WL images is not typically required as part of routine QA. Optimisation of the DeepWL code may be possible to improve the evaluation time per image but it is not prohibitively slow for routine QA in the current state. 

Another limitation of the proposed framework is the expertise required to train DL models and model the linac in the 3D modelling software. In terms of training DL models, there are many freely available U-Net based segmentation codes accessible through platforms such as Github \cite{RN65}. These require little modification to train a similar model presented in the current work and thus mitigates much of the required expertise. Whilst 3D modelling may be considered a niche skill, it is becoming more common with the increasing utilisation of 3D printing technologies \cite{RN67, RN69, RN68, RN70} in radiation oncology for applications such as printing of phantoms, bolus, and brachytherapy surface applicators. This suggests that the difficulty in generating the proposed linac model using the method of the current work compared with a traditional Monte Carlo approach is significantly less given that there is very little or no coding expertise required and accurate geometric modelling of the linac components is unnecessary.

\subsection{Suggestions for Future Work}

We have demonstrated a novel method of training DL models for performing routine linac QA, potentially reducing the amount of time required by a physicist to analyse results and improve the robustness and accuracy of such results. In the current work, we demonstrated this technique using a common QA procedure, the WL test, but this technique could also have been applied to other routine QA procedures such as kV and MV image scaling, isocentre verification, and cone-beam CT image analysis.  

However, unlike traditional automated methods for analysing these types of images, a DL-based approach would theoretically be more robust to phantom setup and other confounding variables since it uses not only relative pixel intensities in the image to guide the analysis, but also through knowledge-based analysis as was demonstrated in the current work.

As was previously discussed, the proposed method of generating MV based images using optical raytracing could also be applied to generating synthetic kV images thus expanding the potential applications to synthetic kV and CT images of patients based on 3D humanoid models. This approach allows for customisable synthetic datasets by animating specific geometric and tissue parameters in the 3D human model. 

In future work, we intend to further develop the model of the current work by training on different MLC field sizes, investigating the effect of statistical image noise and including different phantom designs in the training data to improve model generalisability.

\section{Conclusions}

A novel method of generating synthetic EPID WL QA images simulated using optical path-tracing has been demonstrated in the current work. This approach may be beneficial for those without the coding experience required to develop complex radiation transport Monte Carlo simulations of a linac as well as providing a quick and visually accurate method of generating training data with associated labelled masks for training DL models in radiation oncology. 

The synthetic data generated using this approach was used to train a simple U-Net segmentation model, to predict the displacement between the MLC field and BB in measured EPID WL field without transfer learning techniques. The predictions of this model were shown to be statistically similar to a CED approach for three EPID models with evidence to suggest that the DL model may produce more accurate predictions, particularly on the lower resolution AS500 EPID panel. 

The DL-based approach was also shown to produce more robust predictions to outliers, where the BB approached the edge of the MLC defined field for several artificial test cases. The results also indicated that the DL approach produced segmentations of the BB which were more consistent with the “true” circular geometry and this may contribute to more accurate predictions of displacement, but further evidence is required to confirm this.

The DL model for this application was quick to train and computationally feasible for most medical physics departments without the need for dedicated GPU hardware. This approach has the potential to be applied to the automated analysis of multiple routine linac QA tasks utilising the EPID or kV onboard imaging device.

\section{Acknowledgements}

Special thanks are given to Meghan L. Douglass for providing scientific and statistical advice, independently reviewing the manuscript as well as for the many useful discussions about this project.

\bibliography{bibliography2.bib}  

\end{document}